# Evidence for Intrinsic Redshifts in Normal Spiral Galaxies

David G. Russell

Owego Free Academy, Owego, NY 13827 USA

Russeld1@oagw.stier.org

## Abstract

The Tully-Fisher Relationship (TFR) is utilized to identify anomalous redshifts in normal spiral galaxies. Three redshift anomalies are identified in this analysis: (1) Several clusters of galaxies are examined in which late type spirals have significant excess redshifts relative to early type spirals in the same clusters, (2) Galaxies of morphology similar to ScI galaxies are found to have a systematic excess redshift relative to the redshifts expected if the Hubble Constant is 72 km s$^{-1}$ Mpc$^{-1}$, (3) individual galaxies, pairs, and groups are identified which strongly deviate from the predictions of a smooth Hubble flow. These redshift deviations are significantly larger than can be explained by peculiar motions and TFR errors. It is concluded that the redshift anomalies identified in this analysis are consistent with previous claims for large non-cosmological (intrinsic) redshifts.

## 1. Introduction

Empirical evidence has accumulated which indicates that some quasars and other high redshift objects may not be at the large cosmological distances expected from the traditional redshift-distance relation (Arp1987,1998a, 1999; Chu et al 1998; Bell 2002; Lopez-Corredoira&Gutierrez 2002, 2004; Gutierrez & Lopez-Corredoira 2004). The emerging picture is that some quasars may be ejected from active Seyfert galaxies as high redshift objects that evolve to lower redshifts as they age. Recently, Lopez-Corredoira & Gutierrez (2002, 2004) demonstrated that a pair of high z HII galaxies are present in a luminous filament apparently connecting the Seyfert galaxy NGC 7603 to the companion galaxy NGC 7603B which is previously known to have a discordant redshift. It was noted by Arp (1980) that the NGC 7603/NGC 7603B pair is one of a class of objects in which a lower redshift galaxy appears to be physically connected by a luminous bridge to a companion with a higher discordant redshift.

   The apparent associations and connections between high z quasars and lower z Seyfert galaxies are generally dismissed as accidental alignments of foreground galaxies with background objects.  However, the statistical probabilities that the associations are accidental are very small.  For example, Arp (1999) finds that the probability of having 5 of 6 quasars aligned within +/- 15° of the minor axis of NGC 5985 to be only $10^{-8}$ to $10^{-9}$. Arp&Russell (2001) found that the bright radio quasars 3C37 and 3C39 are paired across the centroid of the disturbed galaxy pair NGC 470/474 in an arrangement with a probability of only $2x10^{-9}$ of being accidental. Zhu&Chu (1995) used the angular correlation function to establish that the quasars in the direction of the Virgo Cluster are associated with Virgo Cluster galaxies with a statistical significance of $7.5\sigma$. Despite the low statistical probability that the associations between quasars and low redshift galaxies can be accidental alignments, the hypothesis has not gained widespread support.

   Arp (1987, 1988, 1990, 1994, 1998a,b) has found evidence that late type spiral galaxies (Sbc/Sc) and smaller companion galaxies tend to have a systematic excess redshift relative to larger early type (Sa/Sb) spiral galaxies.  Russell (2002) utilized linear diameters as a test and confirmed this phenomenon.  Moles et al (1994) presented a strong case for interaction between the low redshift spiral galaxy NGC 450 (cz=1863 km $s^{-1}$) and the higher redshift galaxy UGC 807(cz=11,540 km $s^{-1}$).  Arp (1997) presented evidence that the large redshift discrepancies found in some Hickson compact groups do not result from chance alignments of small foreground and background groups.

   The accumulated evidence suggests that redshift anomalies which cannot be accounted for in the traditional distance-redshift relation may exist from high z quasars down to normal low z galaxies.  Apparent physical connections between low z galaxies and higher z quasars or companion galaxies have offered more direct evidence for the existence of genuine redshift anomalies (Arp 1980, 1982, 1987, 1998a, 2003; Lopez-Corredoira&Gutierrez 2002, 2004; Gutierrez&Lopez-Corredoira 2004; Arp, Burbidge&Burbidge 2004).

   It has been proposed that the redshift anomalies result from non-cosmological (intrinsic) redshifts (See Arp 1987, 1998a for reviews).  In this model the intrinsic redshift component is related to the age of a galaxy such that younger galaxies have a larger intrinsic redshift component than older galaxies.  As a galaxy or quasar ages the size of the intrinsic redshift is predicted to decrease. Narlikar&Arp (1993) proposed a variable mass hypothesis without expansion of the universe to explain intrinsic redshifts. Alternatively Bell (2002) proposed that the intrinsic redshift component is superimposed upon expansion of the universe.

If galaxies and quasars do contain a component of intrinsic redshift then the implications for cosmological models are significant. It is therefore important to establish whether or not the redshift anomalies are real by direct means. Popowski&Weinzierl (2004) have proposed that if quasars are nearby, then they should have detectable proper motions. For normal galaxies, it may be possible to establish the existence of redshift anomalies by determining distances to individual galaxies and groups. For example, it was pointed out by Arp (1988) that the ScI type spirals in Virgo have very large excess redshifts that require large positive peculiar motions in traditional views. However, Sandage (1993) proposed that the large ScI's in Virgo actually have distances of ~28 Mpc. With the completion of the Hubble Key Project (Freedman et al 2001), Cepheid distances are available to three of the ScI's in Virgo and confirm that the ScI's are at distances of 14.9 to 15.9 Mpc. Arp (2002) has pointed out that the new Cepheid distances require that the ScI's in Virgo must have very large and systematic positive peculiar motions or alternatively an intrinsic redshift component.

The Tully-Fisher relation (Tully&Fisher 1977) is the most important secondary distance indicator for determining distances to large numbers of spiral galaxies. The basis of the TFR is the direct relationship that exists between the rotational velocity and the absolute magnitude of spiral galaxies. The general form of the TFR is:

$$M = ax + b \qquad\qquad\qquad\qquad (1)$$

Where M is the absolute magnitude of the galaxy, a is the slope of the relationship, x is the rotational velocity parameter, and b is the zero point. In this analysis the TFR will be utilized to analyze: (1) whether or not late type spiral galaxies in clusters have an excess redshift component relative to early type spirals (Arp 1988), (2) whether or not ScI type spirals have a systematic excess redshift component (Arp 1990), and (3) whether or not galaxies with good TFR distances have large deviations from the Hubble relation that cannot be accounted for by peculiar motions and TFR distance errors.

The paper is organized as follows:  Section 2 discusses calibration of the TFR.  Section 3 considers redshift anomalies in clusters.  Section 4 considers the redshift-distance relation of ScI galaxies.  Section 5 presents examples of galaxies and pairs/groups with extreme deviations from the Hubble relation.  Section 6 considers whether or not Malmquist bias could account for the results.  Section 7 discusses possible explanations for these results and several implications.

## 2.  Calibration of the Tully-Fisher Relationship

The Tully-Fisher relation (TFR – Tully&Fisher 1977) is the most widely used secondary distance indicator for spiral galaxies.  Several recent large TFR studies have used cluster samples to determine the value of the Hubble Constant finding values of $H_0$=72 km s$^{-1}$ Mpc$^{-1}$ (Sakai et al 2000, Freedman et al 2001) and $H_0$=77 km s$^{-1}$ Mpc$^{-1}$ (Tully&Pierce 2000).  Calibration of the TFR may be accomplished with a combination of cluster templates for defining the slope and the use of Cepheid calibrators to determine the zero point (Giovanelli et el 1997a; Sakai et al 2000, Tully&Pierce 2000).  However, the TFR has also been calibrated from field galaxies with absolute magnitudes determined by Hubble distances (Kannappan et al 2002).

Several factors have been identified as important contributors to TFR scatter including morphological type (Roberts 1978; Rubin et al 1985; Giraud 1986; Giovanelli et al 1997a, Corteau et al 2003; Sandage 1996, 1999a,b; Theureau et al 1997; Russell 2002), B-R color (Kappannan et al 2002) and baryonic mass (McGaugh et al 2000).  Russell (2002) demonstrated that within the calibrator sample galaxies of morphology similar to ScI's and Seyfert galaxies are more luminous at a given rotational velocity than Sab/Sb galaxies and Sc galaxies of luminosity classes II to IV.  Russell (2004) demonstrated that this phenomenon is also exhibited by galaxies in cluster samples.

Figure 1 is a plot of the TFR in the B-band.  Open circles are galaxies of morphological types SbcI, SbcI-II, ScI, ScI-II, and Seyfert galaxies (hereafter ScI group).  Filled circles are Sab/Sb/Sbc/Sc galaxies not in the ScI group and generally of luminosity classes II to IV (hereafter Sb/ScIII group).  It is clearly seen in Fig. 1 that the ScI group galaxies are systematically more luminous at a given rotational velocity than Sb/ScIII group galaxies. This systematic difference is corrected for by creating separate type dependent TFR equations for the ScI group and Sb/ScIII group (see also Russell 2002,2004).

Table I lists the twenty-seven calibrators used for calibrating the TFR in this analysis. Column 1 is the galaxy name, column 2 is the morphological type in the LEDA database, column 3 is the rotational velocity from Russell (2002), column for is the corrected B-band magnitude from the LEDA database, column 5 is the calibrator distance, column 6 is the reference for the calibrator distance, column 7 is the individual calibrator zero points using slopes derived from least squares fits to the individual Type Dependent calibrator samples, column 8 gives the individual zero points derived with a slope of 5.84 which is found from a least squares fit to all 27 calibrators, column 9 gives the individual zero points derived with a slope of 5.09 which is the mean slope of the ScI unbiased cluster templates in Russell (2004).

One procedure for calibrating the TFR is to use the calibrator sample alone. In this case the 27 calibrators are split into a group of 12 ScI group galaxies and 15 Sb/ScIII group galaxies. A least squares line was fit to the samples to determine the slope (Fig. 1) and individual zero points were calculated from the slopes resulting in the Type Dependent Tully-Fisher relations (hereafter TD-TFR):

m-M = 19.85(+/- 0.16) + btc + 5.25(+/-0.10)(log Vrot –2.2)   (Sb/ScIII group)     (2)

m-M = 20.46(+/- 0.09) + btc + 4.80(+/-0.16)(log Vrot – 2.2)  (ScI group)             (3)

where m-M is the distance modulus, btc is the corrected B-band magnitude from the LEDA database, and log Vrot is the logarithm of the rotational velocity. Uncertainty in zero point is $1\sigma$ scatter around the  mean zero point and is equivalent to the  empirical scatter around the calibrator distance moduli. The $1\sigma$ scatter of the slope is derived from a Jackknife error analysis (Feigelson&Babu 1992). Note that the primary difference between the two TD-TFR equations is a zero point offset of 0.61 mag. If a single calibration is used, ScI group galaxies will have systematically underestimated TFR distances whereas Sb/ScIII group galaxies will have systematically overestimated TFR distances (see also Russell 2004).

Several important points must be made regarding this calibration. First, the LEDA team does not homogenize corrected data into their database as is commonly thought. The LEDA team collects raw data from the primary sources which is homogenized to a common standard and then corrected for inclination, absorption, and K-effect (Theureau 2004, private communication). Second, it might be objected that the two type dependent calibrator samples are too small for reliable TFR calibration. However, Tully&Pierce (2000) found that the cluster templates provided slope and intercept values virtually identical to those found from the calibrator sample alone in B,R,I, and K' bands. Sakai et al (2000) also found that the I-band TFR from cluster templates was virtually identical to that found from the calibrator sample alone. These results reinforce that useful Tully-Fisher relations can be derived from the calibrator samples.

The scatter of the TD-TFR relative to the calibrator distances is only +/- 0.16 mag (Sb/ScIII group) or +/- 0.09 mag (ScI group). Russell (2004) pointed to reasons why this small scatter is not simply a result of small number statistics. However, an alternative approach would be to increase the sample size by utilizing the entire calibrator sample for determining the slope of the relation and then calculating individual type specific zero points. A least squares line was fit to the entire sample of 27 calibrators which provided a slope of 5.84 and the following TD-TFR equations:

m-M = 19.83(+/- 0.18) + btc + 5.84(log Vrot –2.2)   (Sb/ScIII group)        (4)

m-M = 20.35(+/- 0.11) + btc + 5.84(log Vrot – 2.2)  (ScI group)             (5)

Note that by increasing the slope to 5.84, the mean zero points are reduced by 0.03 mag for the Sb/ScIII group and 0.11 mag for the ScI group. However, it was shown by Russell (2004) that an ScI bias is present in calibrator samples or cluster templates that results in TFR slopes too steep when all galaxies are utilized for slope determination without accounting for morphological type.

A third way to calibrate the TD-TFR is to utilize the slope derived from ScI unbiased cluster templates.  The ScI unbiased TFR slopes of the clusters A2634, A2197/99, Antlia, and NGC 383/507 have a small range of 5.01 to 5.16 (Russell 2004).  Adopting the mean slope of 5.09 gives the equations:

m-M = 19.86(+/- 0.16) + btc + 5.09(log Vrot –2.2)   (Sb/ScIII group)        (6)

m-M = 20.43(+/- 0.09) + btc + 5.09(log Vrot – 2.2)  (ScI group)                (7)

Equations 6 and 7 give distance moduli within +/- 0.09 mag of equations 2 and 3 across the full rotational velocity range of the calibrator sample.  Throughout this paper TD-TFR distances will be calculated with equations 6 and 7 unless otherwise indicated.

## 3.  Dependence of Radial Velocity on Morphology in Large Clusters

### 3.1 The Virgo Cluster

The Virgo Cluster is the nearest large cluster for which significant redshift anomalies have been identified (Arp 1988, Russell 2002).  Arp (1988) noted that the redshift distribution in the Virgo Cluster is morphologically dependent such that late type spirals(Sbc/Sc) have larger mean redshift than early type spirals(Sa/Sb).  ScI galaxies were shown to have very large excess redshifts and Sandage (1993) argued that the large ScI's were at a distance of ~28 Mpc.  However, the Cepheid distances to the large ScI's in Virgo are 14.9 Mpc (NGC 4536), 15.2 Mpc (NGC 4321) and 15.8 Mpc (NGC 4535) (Freedman et al 2001) which confirms that the large excess redshifts are real (Arp 2002).

In order to evaluate the relationship between mean redshift and morphology in the Virgo Cluster, the LEDA database was searched for all galaxies in Virgo with B-band magnitudes and rotational velocities derived from HI linewidths or optical rotation curves.   The sample was restricted to morphological types SO-a to Sc, mean surface brightness in the $D_{25}$ isophote (bri25 in LEDA) from 22.40 to 24.40 mag arc sec$^{-2}$, and log Vrot > 1.89.  The sample was supplemented with E and SO galaxies from the Surface Brightness Fluctuation Method study of Tonry et al (2001).

In Table II, the mean distances and redshift velocities of subsamples of Virgo Cluster galaxies are provided. While all subsamples have mean distances that agree within 1.3 Mpc of the overall mean cluster distance, two outstanding anomalies emerge in Table II. First, there are dramatic differences in mean redshift that are dependent upon morphological type. Second, galaxies with linear diameters <20 kpc within the E and SO group and galaxies with log Vrot < 2.150 in the SO-a to Sb group have significantly larger mean redshifts than the larger galaxies within the same morphological groups.

The most dramatic result in Table II is the extreme excess redshift of the ScI/Seyfert group relative to the large Sab/Sb galaxies in Virgo. The individual galaxies are listed in Table III. *If the difference results from peculiar motions in a strict velocity interpretation of the redshifts, then the ScI's in Virgo are systematically receding from the Milky Way with a mean peculiar velocity of +808 km s$^{-1}$ while the large Sb galaxies in Virgo are systematically approaching the Milky Way with a mean peculiar velocity of –868 km s$^{-1}$.* A Kolmgorov-Smirnov test (K-S) gives P < .001 that the ScI galaxies have the same redshift distribution as the E/SO galaxies in Virgo with SBF distances.

The second discrepant result in Table II is the excess redshift of small E/SO and slow rotating SO-a/Sb galaxies relative to the large galaxies in the same groups. Relative to the larger galaxies in each group, the small galaxies have a mean excess redshift of +417 km s$^{-1}$ for the E/SO group and +837 km s$^{-1}$ for the SO-a/Sb group. This excess extends to the dwarf E/SO galaxies in Virgo (Binggelli, Popescu, & Tammann 1993). Similar redshift excesses for smaller companions have also been demonstrated for the Local Group and M-81 group (Arp 1988, 1994).

### 3.2 Pisces Filament

Russell (2004) selected 30 galaxies from the Pisces filament sample of Giovanelli et al (1997b). Galaxies from the G97 sample were restricted to morphological types Sa to Scd, bri25 from 22.40 to 24.40, log Vrot > 2.00, and inclination > 30°. Table IV shows that while the Sab/Sb and Sbc/Sc galaxies in Pisces have nearly identical mean distance moduli, the mean redshift of the Sbc/Sc galaxies is +552 km s$^{-1}$ greater than the mean redshift of the Sab/Sb galaxies. A K-S test gives P=.001 that the redshift distribution of the Sab/Sb and Sbc/Sc is the same. As found in the Virgo Cluster, the Sbc/Sc galaxies in Pisces have a systematic excess redshift relative to the Sab/Sb galaxies.

### 3.3 Serpens Filament

The LEDA database was searched for galaxies adjacent to the Hercules Supercluster with R.A. coordinates between 15h 25min and 16h 00min and Dec. between +17° to +26° that have rotational velocity and B-band magnitude data. The selected galaxies were restricted to galaxies with morphological types Sa to Sd, bri25 from 22.40 to 24.40, and log Vrot > 2.00. Twenty galaxies were identified with TD-TFR distance moduli from 35.48 to 36.48.

Table IV summarizes the mean data for the 20 galaxies in the Serpens filament and Table V lists the individual galaxies. As found in the Virgo Cluster and Pisces Filament, the Sbc/Sc galaxies have the same mean distance moduli as the Sa/Sb galaxies but a +1604 km s$^{-1}$ larger mean redshift. In addition, there are individual galaxies in the Serpens Filament that have extreme deviations from the mean cluster redshift. The SBbc galaxy PGC 55724 has a redshift of 15066 km s$^{-1}$ which is +4230 km s$^{-1}$ greater than the filament mean. In contrast, the central Sa galaxy PGC 55828 has a redshift of 6989 km s$^{-1}$ which is 3846 km s$^{-1}$ less than the filament mean.

In a strict velocity interpretation, PGC 55828 is a foreground galaxy and PGC 55724 is a background galaxy. However, the distance moduli of both galaxies are within 0.20 mag of the Serpens filament mean distance modulus. So both galaxies would be required to have incorrect TF distances that by chance match the mean distance of the filament. Both PGC 55828 and PGC 55724 exhibit normal double horned hydrogen profiles in Freudling, Haynes, & Giovanelli (1992) – which decreases the likelyhood that large TF errors are the reason for these redshift discrepancies.

In each of these clusters, the Sbc/Sc galaxies have a significant systematic excess redshift relative to Sab/Sb galaxies. Possible explanations and implications of the results found here will be considered in section 7.

## 4. ScI Galaxies

The results of the previous section indicate that late type Sbc/Sc spirals have a systematic excess redshift relative to early type spirals in clusters. Arp (1990) found evidence that this phenomenon extends to field galaxies and Russell (2002) found evidence for the same using galaxy linear diameters.

The sample of Mathewson&Ford (1996 – hereafter MF96) and Giovanelli et al (1997b – hereafter G97) were searched for all ScI group galaxies with corrected B-band magnitudes in the LEDA database. Galaxies were selected for this sample using the following criteria:

(i) Galaxies classified as SbcI, SbcI-II, ScI, ScI-II, or ScII.

(ii) Mean Surface Brightness in the $D_{25}$ isophote (bri25) between 22.40 and 23.99 which is consistent with the calibrator bri25 range.

(iii) Sum of absorption correction to B-band magnitudes from internal absorption (ai) and absorption within the Milky Way (ag) less than 1.00 mag.

(iv) Logarithm of the rotational velocity in MF96 or G97 greater than 2.00 (rotational velocity greater than 100 km s$^{-1}$)

(v) Inclination between 30° and 80°

(vi) Inclination in LEDA agrees within +/- 7° of the inclination in MF96 or G97. This restriction was made to reduce the chances of large errors to corrected B-band magnitudes from errors in LEDA inclinations.

There were 92 galaxies in the MF96 sample and 10 galaxies in the G97 samples that met these criteria. The sample includes field galaxies and cluster galaxies. Distances to the 102 ScI group galaxies were calculated with equation 7. Corrected B-band magnitudes were from LEDA and rotational velocities were taken from MF96 or G97.

Figure 2 is a plot of the Hubble relation for the ScI sample which has a systematic excess redshift relative to a smooth Hubble flow with $H_0$=72 km s$^{-1}$ Mpc$^{-1}$. A K-S test was performed which compared the $H_0$ disribution of the ScI sample against the $H_0$ distribution of the 36 SNeIa from Freedman et al (2001). The K-S test gives P< 0.001 that the ScI sample has the same $H_0$ distribution as the SNeIa sample.

The use of the TD-TFR is not the reason for the difference between the redshift and TF distances. *The ScI group TD-TFR zero point is 0.58 mag larger than the Sb/ScIII group zero point. If a single TF calibration was used, the TF distances of the ScI galaxies would be reduced resulting in a larger discrepancy between the redshift and TF distances.*

The MF96 rotational velocities have an internal consistency of 10 km s$^{-1}$ or better (MF96). Russell (2004) found that the difference between MF96 and G97 rotational velocities is small (1$\sigma$ log Vrot = 0.022). Rotational velocities derived from hydrogen linewidths are available for 62 of the 92 ScI galaxies from the MF96 sample. The 1$\sigma$ scatter between log Vrot from hydrogen linewidths and log Vrot from optical rotation curves is only 0.037 which produces a distance modulus uncertainty of only 0.19 mag with the TF slope of 5.09 utilized in this study.

In order to explain the difference between the TD-TFR distances and the Hubble distances, the MF96 rotational velocities would have to be systematically too small. However, for 35 of the 62 MF96 ScI galaxies with HI rotational velocities in LEDA, the MF96 optical rotation curve velocity is greater than the LEDA HI rotational velocity. Verheijen (2001) and Kannappan et al (2002) have pointed out that the maximum velocity of a rotation curve is often greater than the point at which the rotation curve becomes flat and they argue that Vflat is a better estimate of rotational velocity than Vmax. So both a comparison with LEDA HI rotational velocities and theoretical considerations of rotation curve structure indicate that it is very unlikely that the 102 ScI galaxies have TF distances too small because of large rotational velocity errors.

The results of the comparison between TD-TFR distances and H$_0$=72 redshift distances for 102 ScI galaxies confirms that ScI's have an excess redshift relative to the predictions of a smooth Hubble flow. This confirms the findings of Arp (1990) with a sample selected independently of Arp's sample. Since the sample used here is larger and extends to greater distances and higher redshifts, it strengthens the earlier findings.

## 5. ScI galaxies and companions with extreme deviations from the Hubble relation

Previous claims for redshift anomalies in normal galaxies have primarily been statistical in nature in that the empirical evidence is for systematic redshift differences between galaxies of differing morphological types (Arp 1988, 1994, 1998; Russell 2002). The results discussed thus far in this paper have been a similar nature.

Another possible form of evidence for large redshift anomalies is the existence of discrepancies between the TD-TFR distances and the Hubble distances that are too large to be explained by peculiar motions and TF distance errors. For example, in the Serpens filament the galaxy PGC 55724 has a TF distance modulus of 35.87 – in excellent agreement with the mean filament distance modulus of 35.92. However, the redshift of PGC 55724 is 15066 km s$^{-1}$ which is +4064 km s$^{-1}$ greater than predicted at the Serpens filament distance ($H_0$=72). If PGC 55724 is really at its Hubble distance, then its TF distance modulus is 0.73 mag too small. It is proposed here that similar large redshift deviations may provide virtually unambiguous evidence that the redshift anomalies are genuine.

There are a number of galaxies in the ScI group sample that have extremely large differences between the TD-TFR distance and the $H_0$=72 redshift distance. Individual galaxies were selected from the complete sample of 102 ScI galaxies that met at least one of the following criteria:

(1) The galaxy must have a difference between the TD-TFR distance and the redshift distance that requires a peculiar motion that exceeds 1500 km s$^{-1}$. This velocity difference is adopted because the largest peculiar motions in Virgo are less than 1500 km s$^{-1}$ and it is generally accepted that peculiar motions should not exceed 1500 km s$^{-1}$ (Jensen et al 2003, Karachentsev et al 2003).

(2) The galaxy must have a difference between the TD-TFR distance modulus and the $H_0$=72 distance modulus that exceeds 0.60 mag. for the following reasons: The largest distance modulus errors among the calibrators are +0.35 and –0.20 mag. for the Sb/ScIII group and +0.13 and –0.14 mag for the ScI group. The largest difference from mean cluster distance modulus among the galaxies in the sample of Russell (2004) was 0.60 mag. However, overall scatter around mean cluster distance modulus was only +/-0.22 mag and +/- 0.18 mag for the ScI group galaxies. It is also noted that the largest uncertainty in the B-band magnitude of any of the ScI group galaxies in this sample is 0.60 mag. The largest difference between the log Vrot from LEDA and log Vrot from MF96/G97 is 0.116 which leads to a 0.59 mag distance modulus uncertainty with the TF slope adopted in this study.

The above criteria provided 26 ScI galaxies from the sample of 102 with a TD-TFR distance modulus that was at least 0.60 mag less than the $H_0=72$ distance modulus. Twenty-two of these galaxies would require a peculiar velocity greater than 1500 km s$^{-1}$ in order to account for the difference between the redshift and TF distance. The most extreme example is the SBcII galaxy ESO 384-9 which has a TD-TFR distance of 78.0 Mpc and an $H_0=72$ redshift distance of 159.5 Mpc (Vcmb=11487). If the explanation for this extreme discrepancy is peculiar motions, then ESO 384-9 would have a peculiar motion of +5871 km s$^{-1}$. There is no precedent for such a large peculiar motion in standard views. The discrepancy between the Hubble distance modulus and the TD-TFR distance modulus is 1.55 mag. As noted above, Russell (2004) found that TD-TFR distance modulus errors were never larger than +/- 0.60 mag with a $1\sigma$ scatter of only 0.22 mag for cluster samples and less than 0.16 mag for the calibrators. In addition, the actual uncertainty in the B-band magnitude of ESO 384-9 is only +/- 0.26 mag and the uncertainty in log Vrot is only 0.037. These uncertainties result in a potential uncertainty in the distance modulus of ESO 384-9 of only +/-0.32 mag – which is insufficient to account for the 1.55 mag discrepancy between the TD-TFR distance modulus and the $H_0=72$ distance modulus.

ESO 384-9 provides as unambiguous a case for the existence of genuine redshift anomalies as can be expected from an individual galaxy. Other equally compelling examples include ESO 147-5 (+4539 km s$^{-1}$), ESO 52-20 (+3518 km s$^{-1}$), ESO 30-14 (+3426 km s$^{-1}$), ESO 445-27 (+5278 km s$^{-1}$), ESO 254-22 (+5247 km s$^{-1}$). Each of these galaxies has a discrepancy between the TD-TFR and redshift distance moduli that exceeds 1.20 mag. As noted in section 4, the use of the TD-TFR rather than a traditional single TFR cannot explain these discrepancies because ScI distances are even smaller with a single calibration that does not correct for the type effect.

The MF96 sample was searched for companions to the 26 ScI galaxies that deviate strongly from the Hubble relation. Companions were identified as galaxies that had a redshift within +/- 2000 km s$^{-1}$ of the ScI galaxy and with angular separations that place them within 5 Mpc of the ScI galaxy at the TD-TFR distance. These galaxies are summarized in Table VII. Column 1 is the galaxy identification, column 2 is the morphological type from LEDA, column 3 is the logarithm of the rotational velocity from MF96, column 4 is the corrected B-band magnitude from LEDA, column 5 is the TD-TFR distance modulus, column 6 is the uncertainty in the distance modulus derived from errors in B-band magnitude and rotational velocity, column 7 is the radial velocity corrected to the cosmic microwave background in LEDA, column 8 is the $H_0$=72 distance modulus, column 9 is the peculiar velocity required to account for the difference in redshift and TD-TFR distance. Figure 3 is a plot of the TD-TFR distance modulus of the ScI galaxies against the TD-TFR distance modulus of the companions. The solid line represents equal distance moduli for the companions and ScI's and the dashed line indicates a distance modulus error of 0.60 mag. The companions have distance moduli in excellent agreement with the ScI galaxies with a scatter of only 1$\sigma$= +/-0.13 mag.

Figure 4 is a plot of the $H_0$=72 redshift distance modulus of the pairs or groups in Table VII vs. the mean TD-TFR distance modulus. The solid line in Fig 4 represents equal distance moduli for TD-TFR and $H_0$=72 whereas the dashed line represents the $H_0$=72 distance modulus 0.60 mag greater than the TD-TFR distance modulus. In Figure 5, the mean Vcmb of the pairs and groups in Table VII are plotted against the TD-TFR distance. The solid line represents a Hubble Constant of 72 km s$^{-1}$ Mpc$^{-1}$ and the dashed line represents a peculiar velocity of +1500 km s$^{-1}$.

The companions of the ScI's listed in Table VII strongly reinforce that the redshift anomalies are real. For example, ESO 384-9 has a distance modulus that agrees within 0.03 mag of the distance modulus of the companion ESO 383-56. As a pair the difference between the redshift and TD-TFR distance modulus is 1.66 mag with a velocity excess of +6368 km s$^{-1}$. The ESO 297-36 group includes 6 galaxies for which the $H_0$=72 distance modulus is 0.84 mag greater than the TD-TFR distance modulus. The mean excess redshift for this group is +3197 km s$^{-1}$. Since all six galaxies have such excellent agreement in distance modulus the existence of the large velocity anomaly is strongly reinforced.

The results presented in this section strongly indicate the existence of large redshift anomalies that cannot be explained by peculiar motions or TFR distance errors. The interpretation of these results will be considered in section 7.

## 6. Malmquist Bias

A number of studies have explored the impact of Malmquist bias on TF distance calculations (Giraud 1986; Bottinelli et al 1986; Federspiel et al 1994, Sandage 1999a,b; Ekholm et al 1999). Malmquist bias occurs in magnitude limited samples such that at larger distances the more luminous galaxies are increasingly favored and the fainter end of the luminosity function is lost. In TF studies one signature of a biased sample is an increase in mean rotational velocity with increasing sample distance. In Table VI, the ScI group sample is broken into four equal sized distance bins. The mean rotational velocity increases from log Vrot=2.226 for the galaxies closer than 42.5 Mpc to log Vrot = 2.330 for the galaxies greater than 74.5 Mpc distant which indicates the presence of Malmquist bias.

The predicted effects of Malmquist bias include (1) an increase in the value of the Hubble Constant with increasing sample distance (Bottinelli et al 1986; Federspiel et al 1994; Sandage 1999; Giraud 1986,1987) and (2) a decrease in the value of the Hubble Constant with increasing rotational velocity for redshift limited samples (Federspiel et al 1994). Some researchers have claimed Malmquist bias significantly effects the value of the Hubble constant derived from TF studies (Bottinelli et al 1986; Federspiel et al 1994). For example, Bottinelli et al (1988) concluded that there is a significant trend of a rising Hubble Constant with distance. However, a significant trend is not evident in Table VI. The Hubble Constant increases from 85.2 km s$^{-1}$ Mpc$^{-1}$ to 93.4 km s$^{-1}$ Mpc$^{-1}$. Assuming that the <42.5 Mpc distance bin is an unbiased distance limited sample, the effect of Malmquist bias on the mean Hubble Constant value would be 0.08 mag, 0.07 mag, and 0.20 mag in each successive distance bin.

It is not surprising that the effect of Malmquist bias on $H_0$ calculations for the ScI sample is small. In order for the Malmquist bias to have a significant effect, the method of calculating distances must have a large intrinsic scatter. Since the empirical scatter of the ScI group TD-TFR is +/-0.09 mag for the calibrators, +/-0.18 mag for the cluster samples (Russell 2004), and +/-0.13 mag for the galaxies in Table VII, there is not enough scatter in the TD-TFR for a significant effect from Malmquist bias.

Federspiel et al (1994) studied the effect of Malmquist bias on the Mathewson-Ford-Buchhorn sample of 1355 spiral galaxies which forms the basis for the MF96 sample. They concluded that at a given redshift, the value of the Hubble Constant varies with linewidth such that faster rotating galaxies give a lower Hubble Constant than slower rotating spirals. To test for this effect the 55 ScI galaxies from the 102 in our sample with Vcmb between 3000 and 6000 km s$^{-1}$ were selected and broken into four equal sized rotational velocity bins (Table VIII). There is almost no change in $H_0$ with rotational velocity. The galaxies with log Vrot < 2.201 indicate $H_0$=85.2 while the galaxies with log Vrot > 2.317 indicate $H_0$=83.8.

Another form of bias effecting galaxy clusters is cluster population incompleteness bias (CPIB - Teerikorpi 1987; Bottinelli et al 1987; Sandage 2002) - which has the effect of causing an underestimate of cluster distances because the cluster sample is only complete at the bright end of the luminosity function. When CPIB affects a sample, the faster rotators in the cluster should indicate a larger distance than the slower rotators. The Pisces and Serpens filaments do not indicate the expected results from CPIB. For Pisces, the 16 galaxies with log Vrot > 2.250 have a mean distance modulus of 33.68 while the 14 galaxies with log Vrot < 2.250 have a mean distance modulus of 33.70. For Serpens the 11 galaxies with log Vrot > 2.31 have a mean distance modulus of 35.95 while the 9 galaxies with log Vrot < 2.310 have a mean distance modulus of 35.88. In both filaments the faster and slower rotators agree within 0.07 mag in mean distance modulus but the slower rotators give a slightly *larger* distance in Pisces. This is consistent with the findings of Sakai et al (2000) who concluded that CPIB amounts to only about 0.03 to 0.05 mag.

## 7. Discussion & Conclusion

The results of this study support previous claims for large redshift anomalies in normal spiral galaxies (Arp 1988, 1990, 1994; Russell 2002). It was found that within large clusters there is a tendency for late type Sbc/Sc spirals to have significantly larger mean redshifts than early type Sa/Sb spirals.  The most extreme case is the ScI's in Virgo which have a mean redshift ~ 1500 km s$^{-1}$ greater  than the mean redshift of the large Sab/Sb spirals in Virgo.  The excess of Sbc/Sc spirals relative to Sab/Sb spirals in Pisces and Serpens was +552 km s$^{-1}$ and +1604 km s$^{-1}$ respectively.

   Pisces, Serpens, and Virgo were utilized in this study because they had large enough samples (20+ galaxies) with TF data that met the selection criteria for a meaningful comparison.  The Coma cluster sample utilized by Russell (2004) had 19 galaxies.  The 10 Sab/Sb galaxies from this Coma sample have a mean redshift of 7066 km s$^{-1}$ whereas the Sbc/Sc galaxies have a mean redshift of 7238 km s$^{-1}$.  The difference is much smaller but consistent with the results of the other clusters.  It is worth noting that the only ScI galaxy in the Coma sample – UGC 7978 – has the largest redshift of the sample – 8348 km s$^{-1}$ which is 1201 km s$^{-1}$ larger than the cluster mean.

   The Ursa Major cluster sample utilized by Russell (2004) has 5 ScI galaxies which have a mean redshift of 1228 km s$^{-1}$ compared with 1097 km s$^{-1}$ for the remaining 11 spirals in the sample.  As with Coma, this difference is small but consistent with the results of this analysis.  Finally, it should be noted that in the ESO 297-36 group of 6 galaxies (Table VII), the ScII galaxy ESO 297-36 has a redshift over 1000 km s$^{-1}$ greater than the group mean, whereas the Sb galaxy ESO 297-23 has a redshift over 900 km s$^{-1}$ smaller than the group mean.

   These additional groups support the conclusions of section 3.  In section 4 it was shown that ScI galaxies not only have excess redshifts in clusters, but have excess redshifts relative to the predictions with a Hubble Constant of 72 km s$^{-1}$ Mpc$^{-1}$.  In section 5 it was shown that the discrepancies can result in excess redshifts larger than 5000 km s$^{-1}$ or distance modulus differences > 1.20 mag.  It is not clear what phenomenon can explain the redshift anomalies that the empirical evidence of this study indicate exist. The following possibilities are considered:

(1) *Morphological Density Relation*.  Dressler (1980) established that there is a
    relationship between galaxy environment and morphological type.  It is therefore
    possible that substructure in clusters could lead to anomalous redshift patterns.  For

example, the late type spirals are expected to be found in the outer parts of clusters and could be experiencing greater infall. However, this is problematic as an explanation for the excess redshifts of the ScI's in Virgo. Infall might be expected to cause larger peculiar motions, but there is no reason to expect these motions to be systematically positive for the ScI galaxies and systematically negative for the Sab/Sb galaxies. The ScI's in Virgo are on both the front and the backside of the cluster and yet have systematically positive excess redshifts. In addition, the morphological density relation cannot explain the excess redshift of ScI field galaxies or the extreme redshift anomalies discussed in section 5.

(2) *Value of the Hubble Constant*. The ScI sample is consistent with a Hubble Constant of 85 km s$^{-1}$ Mpc$^{-1}$. Interestingly, the Hubble Key Project (Freedman et al 2001) did find H$_0$= 82 km s$^{-1}$ Mpc$^{-1}$ from the Fundamental plane clusters. However, adopting H$_0$=85 would not eliminate the redshift anomalies identified in this paper. For example, ESO 254-22 and ESO 254-12 would still have a TD-TFR distance modulus 0.96 mag greater than the Hubble distance modulus and the redshift excess would still be +3591 km s$^{-1}$. Increasing the value of the Hubble Constant does not eliminate the systematic difference between Sb and ScI galaxies in clusters either.

(3) *Large Scale Flows*. Some studies have demonstrated that large scale streaming motions may exist in the universe (Courteau et al 2000; Dale et al 1999; Mathewson&Ford 1994). The large scale flows are only 200-400 km s$^{-1}$ which is insufficient to explain any of the redshift anomalies found in this analysis.

(4) *Gravitational Redshifting*. Gravitational redshifting has been found in individual galaxies and galaxy clusters (eg Stiavelli&Setti 1993; Cappi 1995; Broadhurst&Scannapieco 2000). In individual galaxies, gravitational redshifts are as large as ~ 50 km s$^{-1}$ in massive galaxies (Stiavelli&Setti 1993) and ~300 km s$^{-1}$ in clusters (Cappi 1995). While gravitational redshifting is a systematic effect, the effects are too small to account for any of the redshift anomalies identified in this analysis.

(5) *Cepheid Calibration*. The systematic excess of ScI group galaxies could indicate a problem with the cepheid calibration of Freedman et al (2001). Allen&Shanks (2004) proposed a much stronger metallicity correction than that used by Freedman et al (2001).  Using a maximum likelyhood analysis and the large metallicity correction, Allen&Shanks find cepheid distances a mean 0.55 mag larger than Freedman et al for galaxies in the ScI group analyzed by both studies.  Recalibrating the ScI group TD-TFR with the Allen&Shanks distances gives:

$$m-M =  btc + 20.84 + 5.60(\log Vrot-2.2) \qquad\qquad (8)$$

The distance modulus of ESO 254-22 would be 34.94 which is still 0.78 mag less than the $H_0=72$ redshift distance modulus and the redshift velocity discrepancy is 3002 km s$^{-1}$.  Thus a correction to the cepheid distance scale will not eliminate the extreme redshift anomalies.  Nor would the Allen&Shanks cepheid scale explain the morphologically dependent redshift anomalies in clusters.

(6) *Real Peculiar Motions*.  Another interpretation of the results of this analysis is that the redshift anomalies represent real peculiar motions.  However, there is no precedent for such large peculiar motions in galaxy clusters or field galaxies.  In addition, large  peculiar motions cannot explain the systematic excess redshift of the ScI group galaxies as there should be large deficits of redshifts for approaching galaxies.

The large redshift anomalies identified in this analysis are a challenge to explain in standard views. Given that conventional explanations are insufficient, there is reason for considering more exotic alternative explanations.  Arp (1998a,b) has proposed that redshift anomalies of the sort identified in this paper result from intrinsic redshifting. The accumulated empirical evidence presented by Arp suggests that the size of intrinsic redshifts is a function of age such that younger objects have larger redshifts than older objects at the same distance. For example it has been found that apparently ejected quasars generally have decreasing redshifts as distance from the proposed parent galaxy increases (Chu et al 1998, Arp 1998, Bell 2002).

If this explanation is valid, then the results of this analysis suggest that late type spirals(Sbc/Sc) are generally younger than early type spirals (Sa/Sb).  This possibility is consistent with current scenarios for secular evolution of spiral galaxies from late type to early type (Zhang 1999, 2003; Pfenniger et al 1996; Pfenniger 1999). There is also some chemical evidence for evolution of spirals from late type to early type (Dutil&Roy 1999).

There are currently only two proposed mechanisms for large intrinsic redshifts. Narlikar&Arp (1993) proposed a variable mass hypothesis in which particle masses grow as an object ages. Moret-Bailly (2003) has proposed that discrete states in quasar redshifts result from CREIL. Both mechanisms are controversial and require further study. However, until a mechanism for intrinsic redshifting is understood, implications for cosmological models remain uncertain. The variable mass hypothesis implies a non-expanding universe (Narlikar&Arp 1993). However, Bell (2002) and Bell&Comeau (2003) argued that intrinsic redshifts are superimposed upon expansion.

There are two areas with cosmological implications where the empirical evidence for intrinsic redshifts has immediate impact. First, if the intrinsic redshifts are real, then they contaminate calculations of the Hubble Constant. After removing the intrinsic component from the measured redshift a lower value of the Hubble Constant is expected. Arp (2002) attempted to remove the intrinsic component by selecting galaxies that are empirically known to have lower excess redshifts and argued for $H_0 = 55$ km s$^{-1}$ Mpc$^{-1}$. Bell et al (2003) have used the presence of discrete states to determine the value of the Hubble Constant. Both the Hubble Key Project SneIa sample and the ScI sample from this analysis indicate intrinsic redshifts are superimposed in discrete amounts upon a true Hubble Constant of 58 km s$^{-1}$ Mpc$^{-1}$.

Second, if intrinsic redshifts exist, then they contaminate cluster velocity dispersions. In addition, the presence of discrete states in redshifts requires that true random peculiar motions must be small or the discrete states would be washed out (Bell et al 2003). This leads to a second important implication of intrinsic redshifts. The dark matter content of galaxy clusters can be estimated from cluster velocity dispersions. However, if intrinsic redshifts contaminate the true velocity dispersion, then cluster mass estimates may be too large.

The results of this analysis strongly suggest that large redshift anomalies exist in normal galaxies which are most likely non-doppler (intrinsic) in nature. While the exact mechanism for intrinsic redshifts is necessarily speculative at this time, the empirical evidence is consistent with the intrinsic component of redshifts being in discrete amounts and age related such that younger objects have a larger component of intrinsic redshift than older objects at the same distance.

Acknowledgements:

This research has made use of the Lyon-Meudon extragalactic Database (LEDA) compiled by the LEDA team at the CRAL-Observatoire de Lyon (France).   I would also like to thank the referee  - Carlos Gutierrez for helpful.

Table I:  Tully-Fisher Calibrators

| 1 | 2 | 3 | 4 | 5 | 6 | 7 | 8 | 9 |
|---|---|---|---|---|---|---|---|---|
| Galaxy | Type | Log Vrot | btc | m-M calibrator | Ref. | ZP | ZP | ZP |
| ScI group | | | | | | Slope = 4.80 | Slope = 5.84 | Slope = 5.09 |
| N253 | SBcII | 2.292 | 6.99 | 27.98 | 1 | 20.55 | 20.45 | 20.52 |
| N1365 | SBbI (SY) | 2.397 | 9.90 | 31.27 | 2 | 20.42 | 20.22 | 20.37 |
| N1425 | SbII | 2.270 | 10.83 | 31.70 | 2 | 20.53 | 20.46 | 20.51 |
| N2903 | SBbcI-II | 2.282 | 8.86 | 29.75 | 3 | 20.50 | 20.41 | 20.47 |
| N3198 | SBcII | 2.184 | 10.22 | 30.70 | 2 | 20.56 | 20.57 | 20.56 |
| N3627 | SBbII(SY) | 2.326 | 8.98 | 30.01 | 2 | 20.43 | 20.29 | 20.39 |
| N4258 | SbbcII-III (SY) | 2.327 | 8.39 | 29.51 | 2 | 20.51 | 20.38 | 20.47 |
| N4321 | SBbcI | 2.338 | 9.79 | 30.91 | 2 | 20.46 | 20.31 | 20.42 |
| N4535 | SBcI-II | 2.270 | 10.35 | 30.99 | 2 | 20.30 | 20.23 | 20.28 |
| N4536 | SBbcI-II | 2.230 | 10.43 | 30.87 | 2 | 20.30 | 20.26 | 20.29 |
| N4603 | SBcI-II | 2.358 | 11.35 | 32.61 | 4 | 20.50 | 20.34 | 20.46 |
| N7331 | SbcI-II | 2.424 | 9.31 | 30.84 | 2 | 20.45 | 20.22 | 20.39 |
| | | | | | | | | |
| Sb/ScIII group | | | | | | Slope = 5.25 | Slope = 5.84 | Slope = 5.09 |
| | | | | | | | | |
| N224 | SbI-II | 2.411 | 3.20 | 24.48 | 2 | 20.17 | 20.05 | 20.21 |
| N2841 | Sb | 2.488 | 9.52 | 30.74 | 5 | 19.71 | 19.54 | 19.75 |
| N3031 | SabI-II | 2.375 | 7.14 | 27.80 | 2 | 19.74 | 19.64 | 19.77 |
| N3351 | SBbII | 2.232 | 10.06 | 30.00 | 2 | 19.77 | 19.75 | 19.78 |
| N3368 | SBabII | 2.336 | 9.74 | 30.11 | 2 | 19.66 | 19.58 | 19.68 |
| N4527 | SBbcII | 2.260 | 10.63 | 30.75 | 6 | 19.81 | 19.77 | 19.81 |
| N4548 | SBbI-II | 2.279 | 10.66 | 31.05 | 2 | 19.98 | 19.93 | 19.99 |
| N4639 | SBbcII-III | 2.215 | 11.80 | 31.71 | 2 | 19.83 | 19.82 | 19.83 |
| N4725 | SbabI-II | 2.325 | 9.75 | 30.46 | 2 | 20.05 | 19.98 | 20.07 |
| N598 | ScII-III | 2.011 | 5.73 | 24.62 | 2 | 19.88 | 19.99 | 19.85 |
| N2090 | ScII-III | 2.146 | 10.96 | 30.35 | 2 | 19.67 | 19.71 | 19.66 |
| N2403 | ScII-III | 2.104 | 8.24 | 27.54 | 2 | 19.80 | 19.86 | 19.79 |
| N2541 | SBcII-III | 1.989 | 11.61 | 30.25 | 2 | 19.75 | 19.87 | 19.71 |
| N3319 | SBcII-III | 2.043 | 11.34 | 30.62 | 2 | 20.10 | 20.20 | 20.08 |
| N4414 | ScII-III | 2.344 | 10.65 | 31.24 | 2 | 19.83 | 19.75 | 19.86 |

References: 1 – Karachentsev et al (2003),  2- Freedman et al (2001), 3- Drozdovsky&karachentsev (2000) , 4-(Newman et al 1999), 5-Macri et al (2001), 6- Saha et al (2001)

Table II:  Virgo Cluster Hubble Constant

| Type | Sample size | Mpc | Vvir | $H_0$ |
|------|-------------|-----|------|-------|
| Elliptical | 27 | 17.1 | 1062 | 62.1 |
| >20 kpc | 9 | 16.8 | 784 | 46.7 |
| <20 kpc | 18 | 17.2 | 1201 | 69.8 |
| | | | | |
| SO-a to Sb | 15 | 17.8 | 913 | 51.3 |
| Log V > 2.150 | 9 | 18.2 | 578 | 31.8 |
| Log V< 2.150 | 6 | 17.3 | 1415 | 87.8 |
| | | | | |
| Sbc/ScII-III to IV | 23 | 19.3 | 1268 | 65.7 |
| | | | | |
| Sbc and Sc I/I-II, Seyferts | 11 | 18.6 | 1855 | 99.7 |
| | | | | |
| All | 76 | 18.1 | 1210 | 66.9 |

Table III  Peculiar Velocities of Sab/Sb and ScI in Virgo

| 1 | 2 | 3 | 4 | 5 |
|---|---|---|---|---|
| Galaxy | Type | Mpc | Vvir | $PV_{72}$ |
| ScI Group | | | | |
| N4254 | ScI.3 | 18.8 | 2508 | +1154 |
| N4303 | SBbcI | 16.3 | 1619 | +445 |
| N4321 | SBbcI | 15.2 | 1689 | +595 |
| N4501 | SbI-II(Sy) | 19.3 | 2379 | +989 |
| N4535 | SBcI-II | 15.8 | 2029 | +891 |
| N4536 | SBbcI-II | 14.9 | 1846 | +773 |
| N4388 | SbII-III (Sy) | 20.8 | 2608 | +1110 |
| N4450 | SabII-III(Sy) | 20.4 | 2068 | +599 |
| N4579 | SBbII(Sy) | 23.6 | 1607 | -92 |
| | | | | |
| Sab/Sb Group | | | | |
| N4192 | SBabII | 15.1 | -46 | -1133 |
| N4216 | SBbII | 14.9 | 219 | -854 |
| N4343 | SbIII | 28.1 | 1074 | -949 |
| N4380 | SbI-II | 23.6 | 1047 | -652 |
| N4402 | SbIV | 19.3 | 328 | -1062 |
| N4548 | SBbI-II | 16.2 | 589 | -577 |
| N4569 | SbabI-II | 12.6 | -137 | -1044 |
| N4698 | SabII | 24.4 | 1083 | -674 |

Table IV: Pisces and Serpens Redshift-Morphology Dependence

| 1 | 2 | 3 | 4 | 5 | 6 | 7 |
|---|---|---|---|---|---|---|
| Cluster | Type | n | m-M | Mpc | Vcmb | H$_0$ |
| Pisces | Sa/Sb | 12 | 33.70 | 55.0 | 4418 | 80.3 |
| Pisces | Sbc/Sc | 18 | 33.68 | 54.5 | 4970 | 91.2 |
| Pisces | All | 30 | 33.69 | 54.7 | 4749 | 86.8 |
| | | | | | | |
| Serpens | Sa/Sb | 8 | 35.92 | 152.8 | 9873 | 64.6 |
| Serpens | Sbc/Sc | 12 | 35.92 | 152.8 | 11477 | 75.1 |
| Serpens | All | 20 | 35.92 | 152.8 | 10835 | 70.9 |

Table V: Serpens Filament

| Galaxy | Type | RA | Dec | logVrot | Incl. | m-M(B) | Vcmb |
|---|---|---|---|---|---|---|---|
| 55198 | Sab | 15.480 | 25.748 | 2.342 | 75 | 35.79 | 10365 |
| 55618 | SBb | 15.618 | 25.565 | 2.459 | 55 | 36.12 | 10562 |
| 55724 | SBbc | 15.662 | 25.742 | 2.487 | 63 | 35.87 | 15066 |
| 55810 | Sc | 15.695 | 23.204 | 2.388 | 46 | 36.23 | 10477 |
| 55828 | Sa | 15.705 | 23.808 | 2.362 | 50 | 36.10 | 6989 |
| 55979 | Sc | 15.762 | 22.879 | 2.371 | 32 | 36.32 | 12378 |
| 56038 | Sc | 15.789 | 21.531 | 2.349 | 53 | 35.94 | 12651 |
| 56175 | Sbc | 15.844 | 20.382 | 2.305 | 55 | 36.21 | 11152 |
| 56227 | Sa | 15.858 | 24.435 | 2.239 | 67 | 35.81 | 9698 |
| 56169 | Sc | 15.841 | 18.139 | 2.507 | 36 | 36.04 | 14095 |
| 55721 | SBab | 15.659 | 23.199 | 2.081 | 56 | 35.98 | 11620 |
| 55708 | Sab | 15.655 | 24.455 | 2.438 | 71 | 35.82 | 10429 |
| 55380 | Sa | 15.550 | 25.569 | 2.221 | 54 | 35.79 | 10177 |
| 55872 | Sc | 15.725 | 17.313 | 2.284 | 50 | 35.59 | 9127 |
| 56532 | Sc | 15.974 | 18.039 | 2.390 | 63 | 35.72 | 12746 |
| 55530 | Sc | 15.594 | 21.503 | 2.200 | 56 | 35.67 | 12697 |
| 86655 | Sc | 15.566 | 21.790 | 2.156 | 68 | 36.48 | 7175 |
| 56186 | Sc | 15.847 | 22.239 | 2.234 | 56 | 35.72 | 9533 |
| 56163 | SBa | 15.835 | 25.055 | 2.301 | 38 | 35.71 | 9609 |
| 55213 | SBbc | 15.484 | 25.458 | 2.371 | 51 | 35.49 | 10163 |

Table VI: ScI sample summary

| 1 | 2 | 3 | 4 | 5 | 6 |
|---|---|---|---|---|---|
| Distance Bin | n | log Vrot | bri25 | i | $H_0$ |
| (Mpc) | | | | | |
| <42.5 | 25 | 2.226 | 23.24 | 58 | 85.2 |
| 44.1-55.2 | 26 | 2.274 | 23.17 | 58 | 88.5 |
| 55.5-73.8 | 26 | 2.253 | 23.22 | 57 | 87.9 |
| >74.4 | 25 | 2.330 | 23.20 | 54 | 93.4 |

Table VII:   ScI galaxies and companions with anomalous redshifts

| 1 | 2 | 3 | 4 | 5 | 6 | 7 | 8 | 9 |
|---|---|---|---|---|---|---|---|---|
| Galaxy | Type | logVrot | btc | $m-M_{TF}$ | +/- | Vcmb | $m-M_{72}$ | $PV_{72}$ |
| 52-20 | SBbcI-II | 2.196 | 13.59 | 34.00 | .18 | 8068 | 35.25 | +3518 |
| 53-2 | Sc | 2.170 | 14.32 | 34.03 | .21 | 7950 | | |
| | | | | | | | | |
| 30-14 | ScI-II | 2.228 | 13.54 | 34.11 | .54 | 8207 | 35.28 | +3426 |
| 30-9 | ScII | 2.491 | 12.90 | 34.24 | .20 | 8113 | | |
| | | | | | | | | |
| 545-13 | ScI-II | 2.350 | 13.52 | 34.71 | .47 | 9923 | 35.70 | +3623 |
| 545-18 | Sbc | 2.276 | 14.55 | 34.79 | .48 | 9770 | | |
| | | | | | | | | |
| 147-5 | SBcI-II | 2.274 | 13.85 | 34.65 | .91 | 10666 | 35.85 | +4539 |
| 147-10 | Sbc | 2.281 | 14.49 | 34.76 | .26 | 10812 | | |
| | | | | | | | | |
| 254-22 | ScII | 2.377 | 13.11 | 34.44 | .35 | 10812 | 35.88 | +5247 |
| 254-12 | Sc | 2.204 | 14.46 | 34.34 | .31 | 9203 | | |
| | | | | | | | | |
| 553-44 | SBcI-II | 2.398 | 13.31 | 34.75 | .19 | 8349 | 35.30 | +2037 |
| 554-1 | Sc | 2.161 | 14.97 | 34.63 | .16 | 8209 | | |
| | | | | | | | | |
| 384-9 | SBcII | 2.260 | 13.72 | 34.46 | .32 | 11487 | 36.01 | +5871 |
| 383-56 | Sb | 2.212 | 14.52 | 34.43 | .16 | 12380 | | |
| | | | | | | | | |
| N3029 | ScI-II | 2.207 | 13.81 | 34.28 | .29 | 6917 | 34.91 | +1747 |
| N2980 | SBcII | 2.387 | 12.81 | 34.19 | .13 | 6063 | | |
| | | | | | | | | |
| 286-79 | SBcI-II | 2.468 | 11.58 | 33.37 | .30 | 4751 | 34.10 | +1353 |
| 235-20 | SBcI-II | 2.188 | 12.76 | 33.13 | .22 | 4515 | | |
| 236-37 | SbcI-II | 2.248 | 12.98 | 33.65 | .27 | 5384 | 34.37 | +1518 |
| 237-2 | SBbcI-II | 2.362 | 12.43 | 33.69 | .46 | 4983 | | |
| | | | | | | | | |
| 119-38 | SBcII | 2.134 | 14.41 | 34.50 | .47 | 8937 | 35.47 | +3220 |
| 120-3 | Sc | 2.204 | 14.70 | 34.58 | .22 | 8355 | | |
| | | | | | | | | |
| 286-58 | SBcI-II | 2.143 | 14.24 | 34.38 | .56 | 9187 | 35.53 | +3774 |
| 286-18 | SBbc | 2.509 | 12.93 | 34.36 | .46 | 9015 | | |
| | | | | | | | | |
| 306-28 | SBcII | 2.130 | 13.88 | 33.96 | .34 | 7254 | 35.02 | +2797 |
| 363-29 | Sb | 2.250 | 14.43 | 34.45 | .42 | 7753 | | |
| 364-8 | Sc | 2.188 | 14.17 | 33.97 | .24 | 7710 | | |
| | | | | | | | | |
| 284-16 | SBcII | 2.188 | 13.20 | 33.57 | .19 | 5282 | 34.33 | +1552 |
| 284-13 | Sb | 2.262 | 14.22 | 34.40 | .15 | 5562 | | |
| 284-21 | SBcIII | 2.111 | 14.33 | 33.74 | .52 | 5655 | | |
| 284-29 | SBbII-III | 2.201 | 13.04 | 32.90 | .20 | 5088 | | |
| | | | | | | | | |
| 469-7 | SBcII | 2.316 | 13.71 | 34.73 | .19 | 8450 | 35.35 | +2092 |
| 406-38 | Sbc | 2.260 | 14.84 | 35.01 | .22 | 8067 | | |
| 406-16 | Sbc | 2.158 | 15.15 | 34.80 | .16 | 8199 | | |
| | | | | | | | | |
| 297-36 | ScII | 2.275 | 14.14 | 34.95 | .23 | 10972 | 35.91 | +3938 |
| 297-23 | Sb | 2.483 | 13.48 | 34.78 | .22 | 9008 | | |
| 297-29 | Sbc | 2.276 | 14.34 | 34.59 | .21 | 9834 | | |
| 354-6 | Sbc | 2.176 | 15.16 | 34.90 | .26 | 9691 | | |
| 354-24 | Sbc | 2.248 | 14.72 | 34.83 | .22 | 10252 | | |
| 354-19 | Sab | 2.444 | 13.95 | 35.05 | .19 | 9732 | | |
| | | | | | | | | |
| 234-22 | SBbcI-II | 2.314 | 12.75 | 33.76 | .15 | 5593 | 34.45 | +1525 |
| 234-13 | SbcII-III | 2.225 | 13.80 | 33.78 | .21 | 4673 | | |
| 234-24 | Sbc | 2.265 | 13.66 | 33.85 | .21 | 5688 | | |
| 234-32 | SBbc | 2.233 | 13.88 | 33.90 | .16 | 5748 | | |
| 186-47 | SBbcI-II | 2.210 | 13.42 | 33.90 | .31 | 4482 | | |
| 186-21 | SBcI | 2.301 | 13.18 | 34.12 | .15 | 5539 | | |
| 234-19 | Sbc | 2.170 | 13.85 | 33.56 | .20 | 5425 | | |
| 234-44 | Sc | 2.124 | 14.46 | 33.93 | .33 | 6341 | | |
| 234-61 | Sc | 2.009 | 14.91 | 33.80 | .22 | 6234 | | |
| | | | | | | | | |
| 501-75 | ScII | 2.250 | 12.54 | 33.22 | .17 | 5563 | 34.44 | +2388 |
| Hydra | Cluster | | | 33.20 | | 4037 | | |

Table VIII: 55 ScI's with Vcmb from 3000 – 6000 km s$^{-1}$

| Log Vrot range | # galaxies | $H_0$ |
|---|---|---|
| <2.201 | 14 | 85.2 |
| 2.210-2.250 | 14 | 91.9 |
| 2.253-2.316 | 14 | 84.4 |
| >2.317 | 13 | 83.8 |

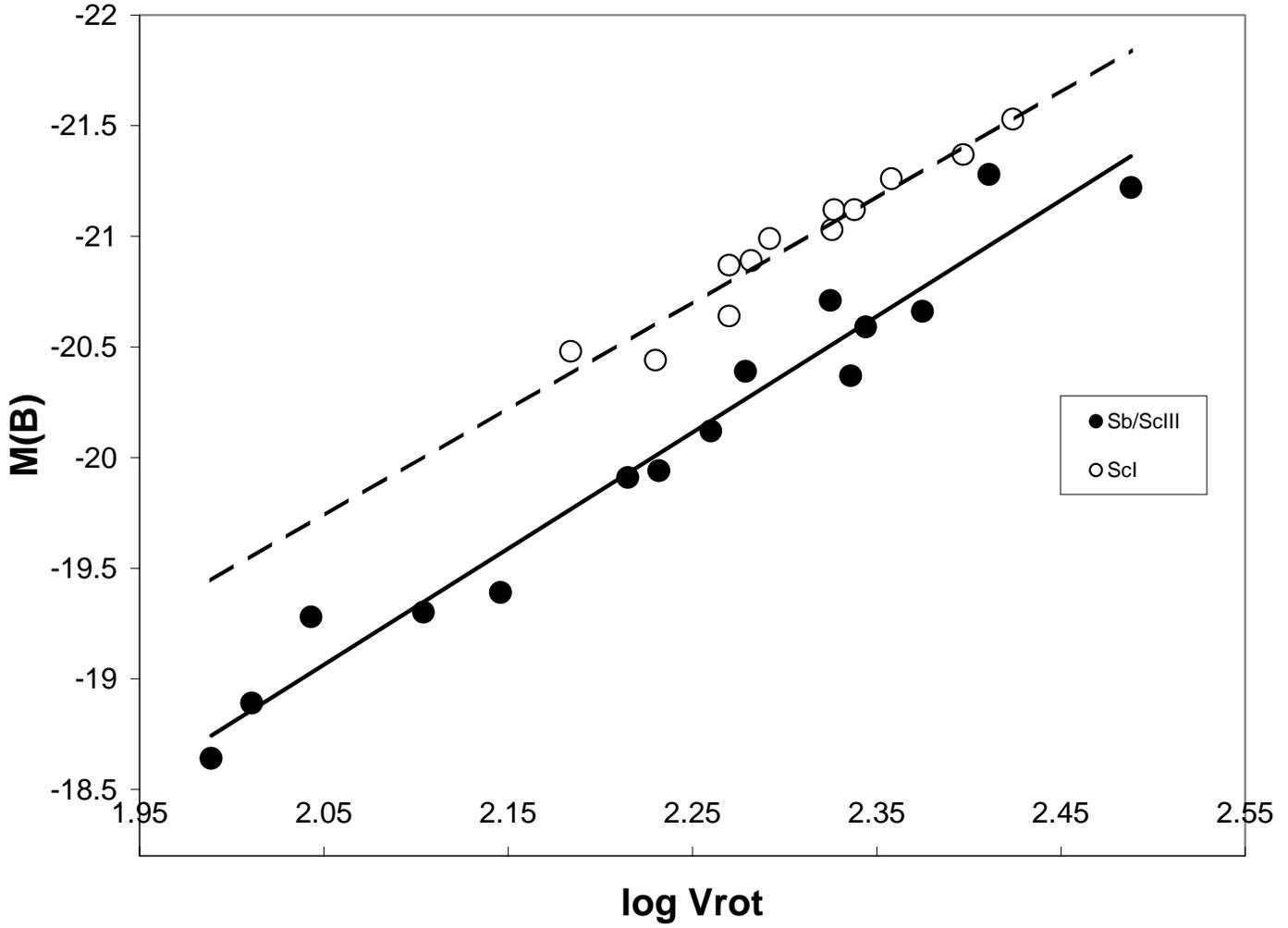

Figure 1: Calibration of the TD-TFR in the B-band. Filled circles are Sb/ScIII group galaxies and solid line is least squares fit. Open Circles are ScI group galaxies and dashed line is least squares fit.

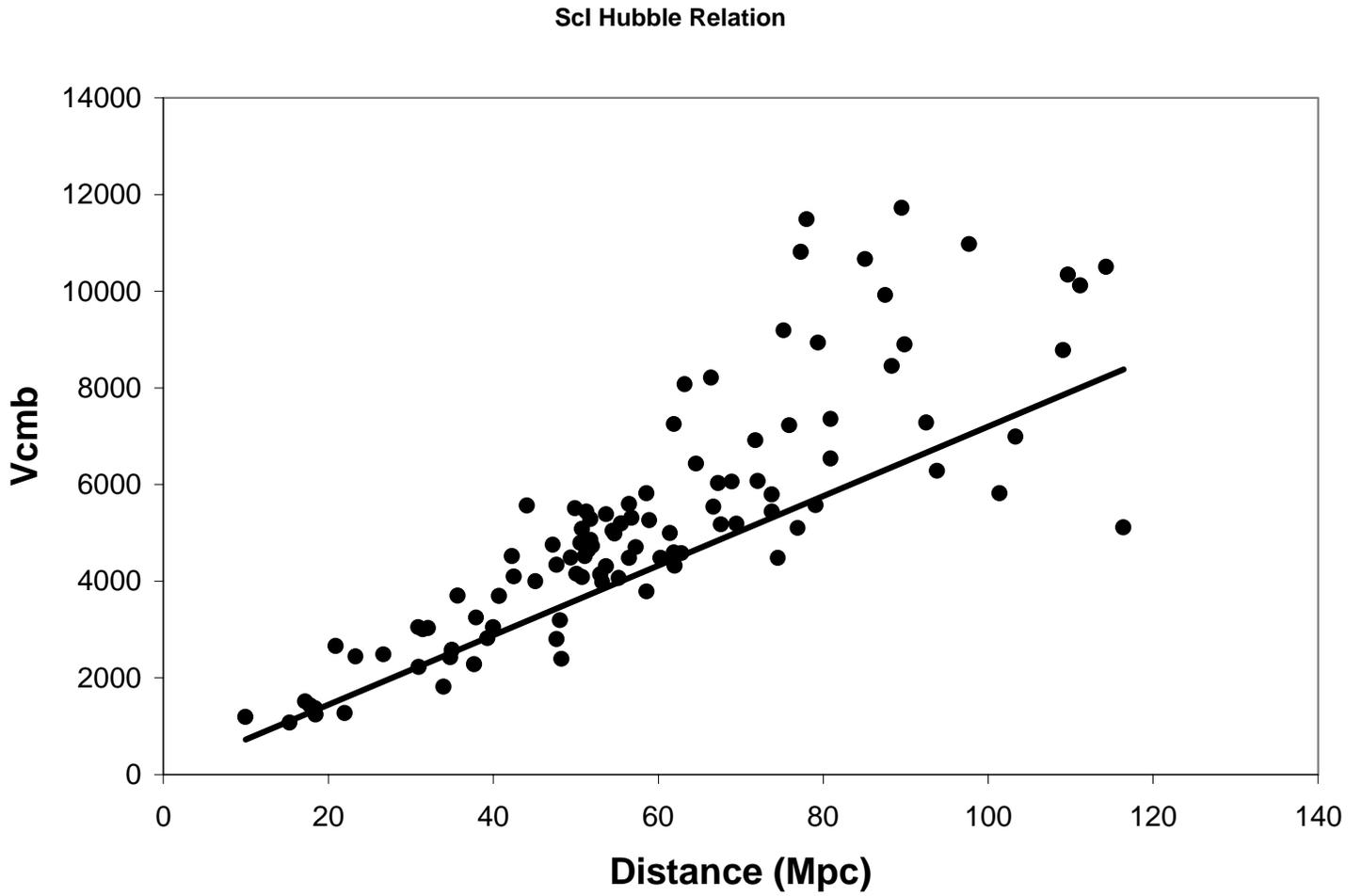

Figure 2 – Hubble relation for 102 ScI group galaxies. Solid line is a Hubble Constant of 72 km s$^{-1}$ Mpc$^{-1}$.

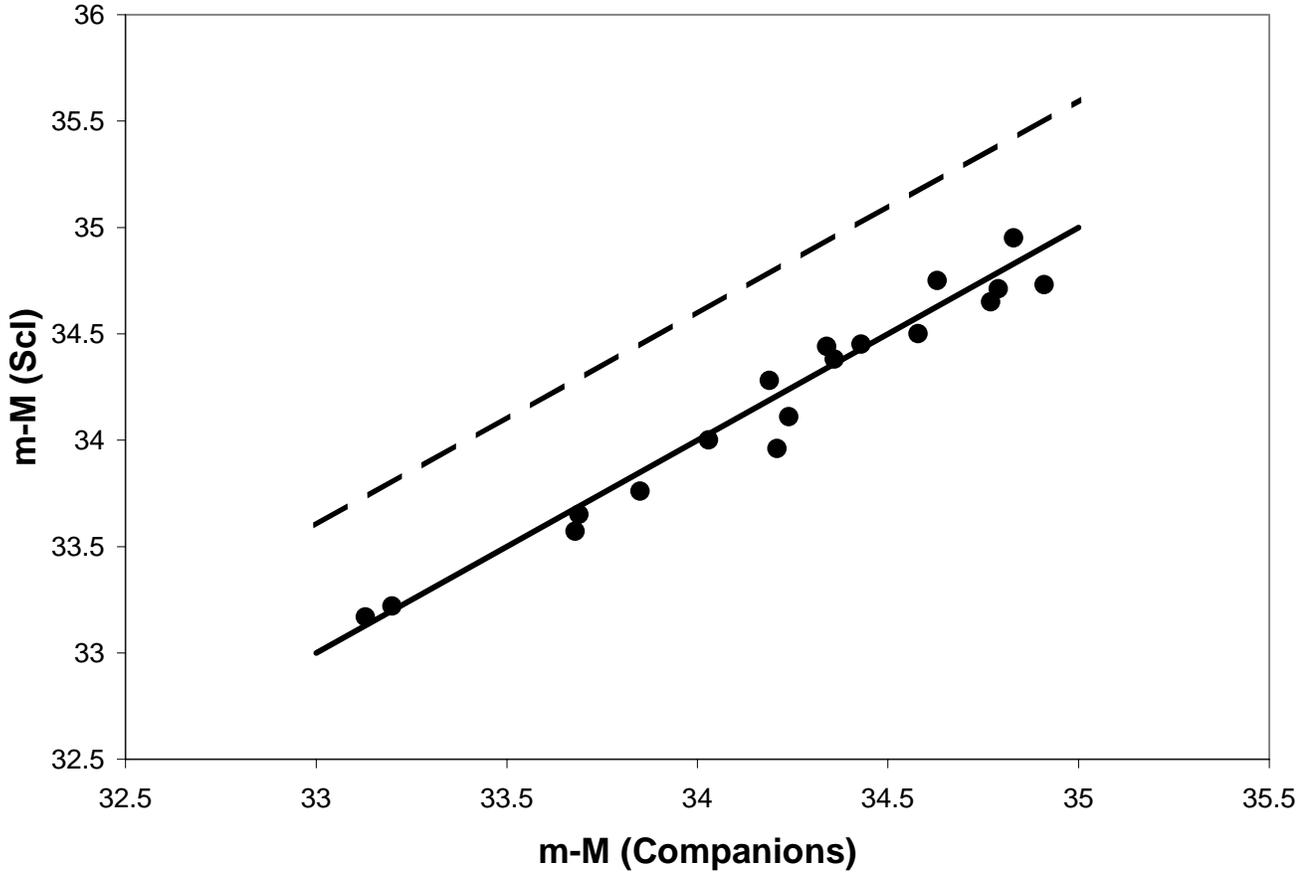

Figure 3: Comparison of TD-TFR distance moduli for ScI galaxies with large anomalous redshifts with the TD-TFR distance moduli for nearby companion galaxies. Solid line represents equal distance moduli for ScI's and companions. Dashed line represents a 0.60 mag distance modulus difference.

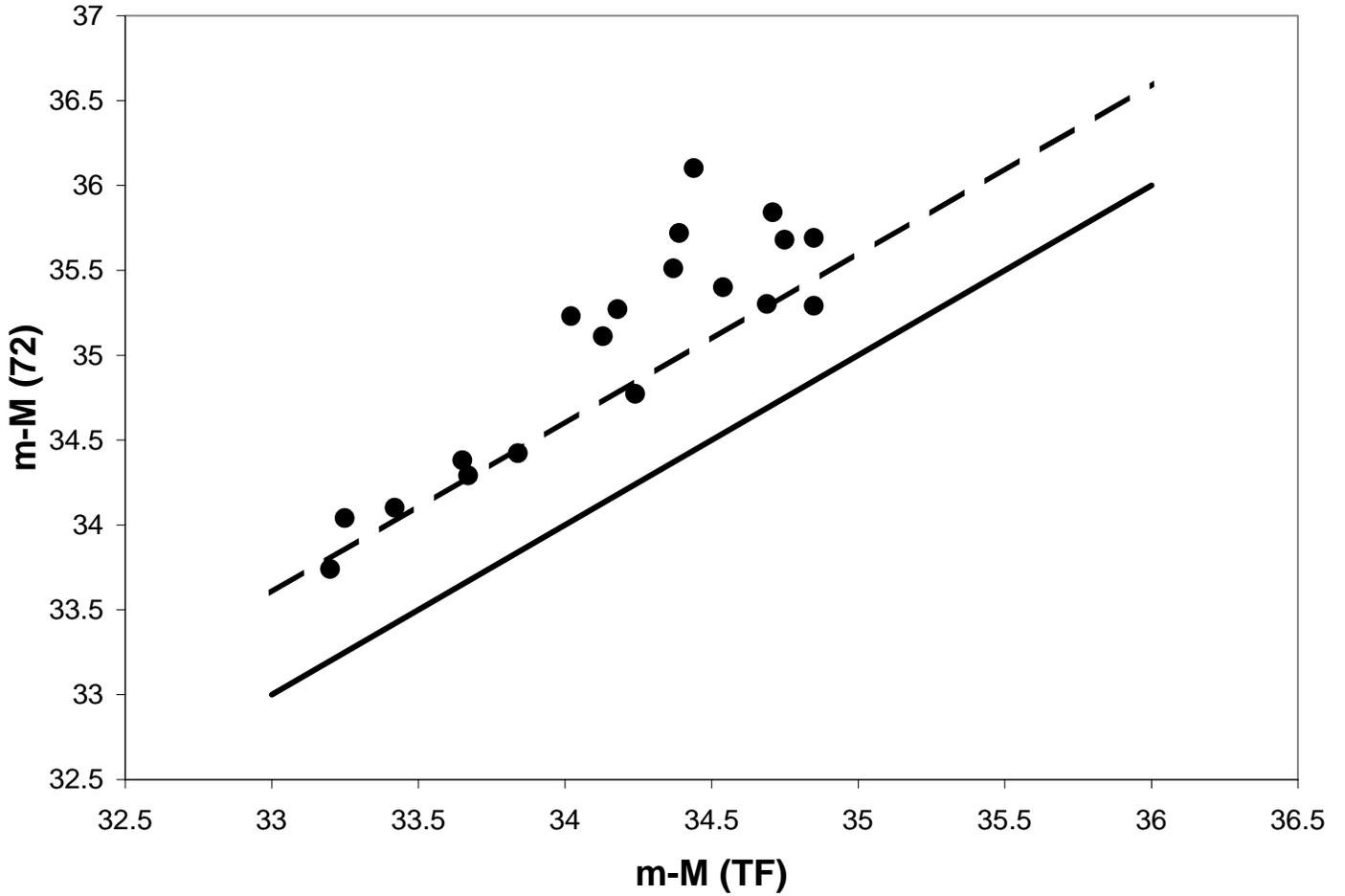

Figure 4: Comparison of $H_0$=72 distance moduli against TD-TFR distance moduli for ScI galaxies with large redshift anomalies. Solid line represents equal distance moduli and dashed line represents redshift distance modulus 0.60 mag greater than TF distance modulus.

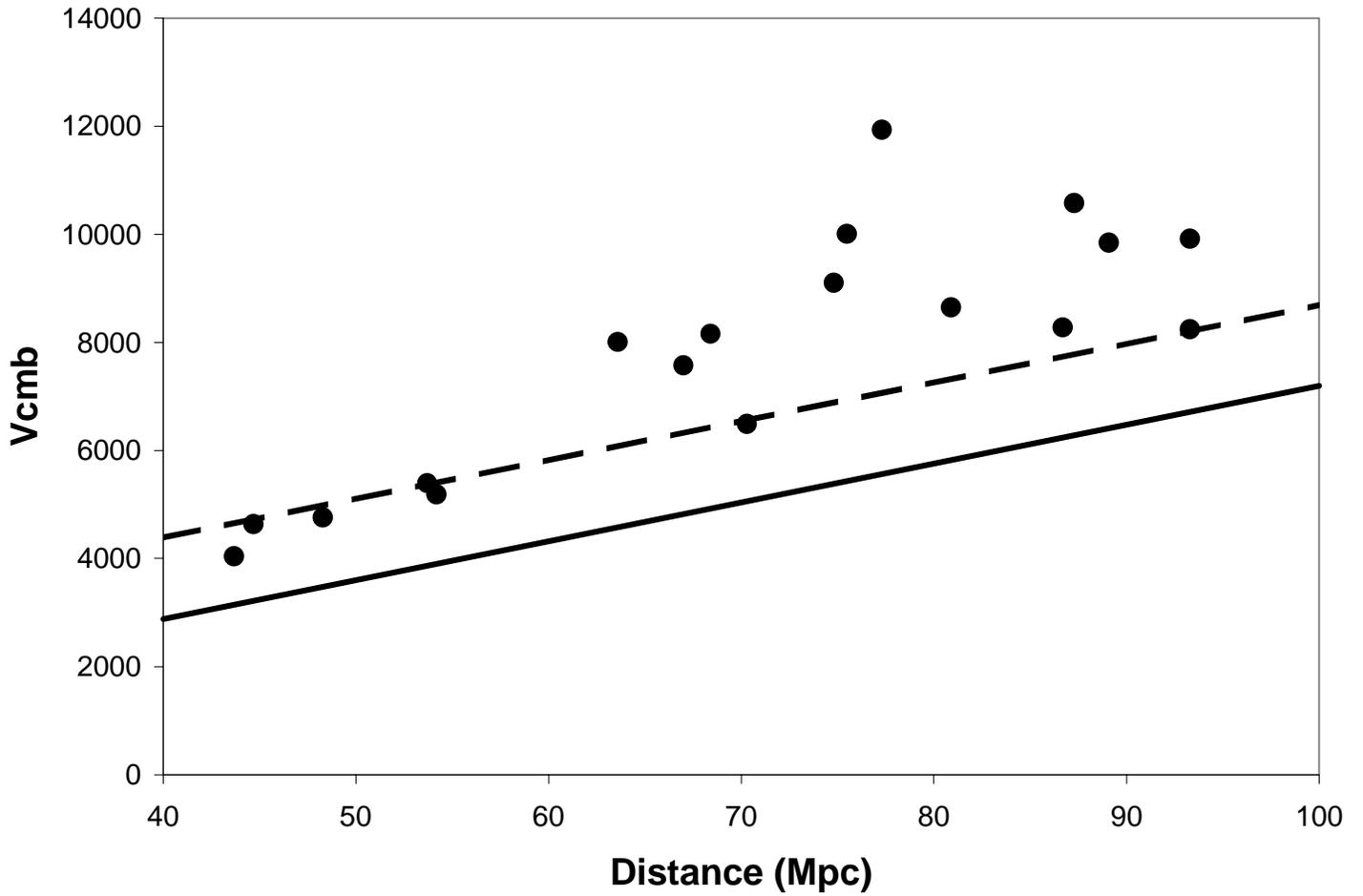

Figure 5 – Hubble plot for ScI galaxies with large anomalous redshifts. Solid line represents a Hubble Constant of 72 km s$^{-1}$ Mpc$^{-1}$. Dashed line represents a peculiar velocity of +1500 km s$^{-1}$.

## APPENDIX:  LEDA Images for ScI galaxies with large excess redshifts

The 5' images from the LEDA database provide strong additional support for the large discrepancy between redshift and Tully-Fisher distances that was found for some of the galaxies in the ScI group sample.  As illustrations, the following compares two of the ScI's which have been argued to have extreme excess redshifts with galaxies within the Serpens filament.  The importance of this comparison is that the galaxies in the Serpens filament give $H_0$=70 km s$^{-1}$ Mpc$^{-1}$ and therefore there is no reason to doubt their TF distances in traditional interpretations.

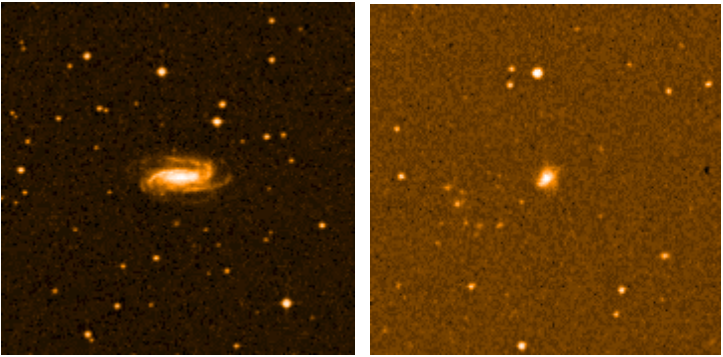

Figure A1 – LEDA 5' images of ESO 445-27 (SBcI-II  -  left image) and PGC 55979 (Sc - right image).  These two galaxies have very similar rotational velocities, mean surface brightness in the $D_{25}$ isophote (bri25 in LEDA), and redshift velocity.  Since the redshifts are close they should be at about the same distance and have nearly the same appearance in the images.  It is clearly seen that ESO 445-27 appears significantly larger with much greater degree of image resolution as expected if it is much closer than PGC 55979 as indicated by TF distances.

| Galaxy | Rotational velocity | bri25 | Mpc TD-TFR | Vcmb | Mpc ($H_0$=72) |
|---|---|---|---|---|---|
| ESO 445-27 | 245 km s$^{-1}$ | 22.87 | 89.5 | 11722 | 162.8 |
| PGC 55979 | 235 km s$^{-1}$ | 23.05 | 183.7 | 12378 | 171.9 |

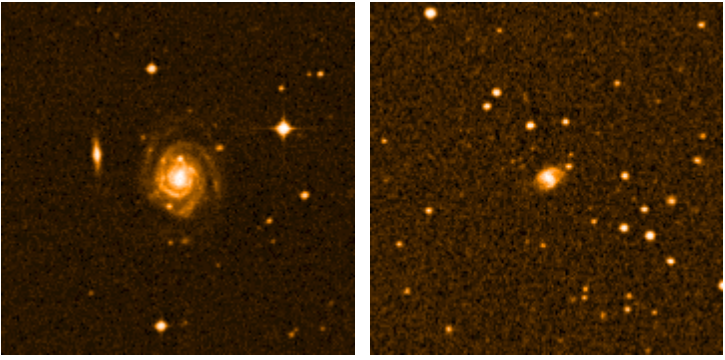

Figure A2 – LEDA 5' images of ESO 147-5 (SBcI-II - left image) and PGC 55872 (Sc - right image). Same as figure A1, but note that ESO 147-5 has an even larger redshift than PGC 55872 and therefore should be even more distant in traditional views. The images support the much closer distance of ESO 147-5 indicated from the TFR and therefore the large implied intrinsic redshift.

| Galaxy | Rotational velocity | bri25 | Mpc TD-TFR | Vcmb | Mpc (H$_0$=72) |
|--------|--------------------|-------|-----------|------|---------------|
| ESO 147-5 | 188 km s$^{-1}$ | 23.22 | 85.1 | 10666 | 148.1 |
| PGC 55872 | 192 km s$^{-1}$ | 22.88 | 131.2 | 9127 | 126.8 |